\documentclass[10pt,conference]{IEEEtran}

\usepackage{amssymb}   
\usepackage{amsfonts} 
\usepackage{amsmath}  
\usepackage{latexsym} 
\usepackage{epsfig} 
\usepackage{psfrag} 
\usepackage{bm} 
\usepackage{xspace} 
\usepackage{graphicx} 
\usepackage{cite} 
\usepackage{url}
\usepackage[latin1]{inputenc}
\renewcommand{\baselinestretch}{1.02}

  {\noindent \emph{Proof:}}{\hfill$\square$} 

\newtheorem{theorem}{Theorem}[section] 
 \newtheorem{lemma}[theorem]{Lemma}

  \newtheorem{definition}[theorem]{Definition} 
  \newtheorem{example}[theorem]{Example} 
  
  \newtheorem{remark}[theorem]{Remark}

\newenvironment{sbmatrix}[1]{\left[\begin{array}{#1}}{\end{array}\right]}

\newcommand{\dfree}{\mbox{$d_{\mbox{\rm\tiny free}}$}}
\newcommand{\F}{\mathbb{F}}
\def\wt{\operatorname{wt}}
\newcommand{\C}{\mathcal{C}}

\renewcommand{\vec}[1]{\ensuremath{\text{\mathversion{bold}$#1$}}}
\newcounter{abc}

\newcommand{\vect}[1]{\mathbf{#1}}

\newenvironment{alphalist}{\begin{list}{(\alph{abc})\hfill}{\usecounter{abc}
     \topsep.5ex \labelwidth.6cm \leftmargin.7cm \labelsep.1cm
     \rightmargin0cm \parsep0ex \itemsep.6ex
     \partopsep1.6ex}}{\end{list}}
\newcommand{\Footnotemark}[1]{${}^{#1}$}
\newcommand{\Footnotetext}[2]{\begin{figure}[!b]\footnotesize%
  \vspace{-3ex}
\makebox[0em]{}\hfill\makebox[0em]{}%
  \par${}^{#1}$ #2\vspace{-0.60ex}\end{figure}\addtocounter{figure}{0}}

\begin{document}

\title{Decoding of MDP Convolutional Codes \\ over the Erasure Channel}

\author{
\IEEEauthorblockN{Virtudes Tom\'{a}s\Footnotemark{*}}
\IEEEauthorblockA{Department of Computational Science\\
and Artificial Intelligence\\
University of Alicante\\Alicante, Spain\\
Email: vtomas@dccia.ua.es
}
\and
\IEEEauthorblockN{Joachim Rosenthal\Footnotemark{\circ}}
\IEEEauthorblockA{Mathematics Institute \\
University of Z\"urich \\
 Winterthurerstr 190\\
CH-8057 Z\"urich, Switzerland\\
www.math.uzh.ch/aa}
\and
\IEEEauthorblockN{Roxana Smarandache\Footnotemark{\ddagger}}
\IEEEauthorblockA{Department of Mathematics and Statistics\\
San Diego State University\\
San Diego, CA 92182-7720, USA\\
Email: rsmarand@sciences.sdsu.edu}
}

\maketitle

\begin{abstract}
  This paper studies the decoding capabilities of maximum
  distance profile (MDP) convolutional codes over the erasure
  channel and compares them with the decoding capabilities of MDS
  block codes over the same channel. The erasure channel
  involving large alphabets is an important practical channel
  model when studying packet transmissions over a network, e.g,
  the Internet.
\end{abstract}

\mbox{} \\[-0.6cm]
\Footnotetext{*}{Partially supported by Spanish grant
MTM2008-06674-C02-01 and a grant of the Vicerectorat
d'Investigaci\'{o}, Desenvolupament i Innovaci\'{o} of the
Universitat d'Alacant for PhD students during a stay at Z\"urich 
Universit\"at on charge to the same program.}
\Footnotetext{\circ}{ Supported by the Swiss National Science
Foundation under Project no. 113251.} 

\Footnotetext{\ddagger}{Supported by
                 NSF Grants DMS-0708033 and TF-0830608.}


\section{Introduction}

When transmitting over an erasure channel like the Internet, one of
the problems encountered is the delay experienced on the received information
which is due to the possible re-transmission of lost packets.  One way
to eliminate these delays is by using forward error correction.

Until now only block codes have been used for such a task, see
\cite{ep58,fa08p}.  In this paper we demonstrate how 
maximum distance profile (MDP) convolutional codes provide
an attractive alternative.

Convolutional codes have a certain flexibility given by the ``sliding
window'' characteristic.  This means that the received information can
be grouped in blocks or windows in many ways, depending on the erasure
bursts, and then be decoded by decoding the ``easy'' blocks
first. This flexibility in grouping information brings certain freedom
in the handling of sequences; we can split the blocks in smaller
windows, we can overlap windows, etc., we can proceed to decode in a
less strict order.  The blocks are not fixed as in the block code
case, i.e., they do not have a fixed grouping of a fixed length.  We can
slide along the transmitted sequence and decide the place where we
want to start our decoding depending on the erasure occurrence.  This
property allows us to correct in a given block more erasures than a
block code of that same length could do.

An $[N,K]$ block code used for transmission over an erasure
channel can correct up to $N-K$ erasures in a
given block.  The optimal error capability of $N-K$ is achieved
by an  $[N,K]$ maximum distance separable (MDS) code.

As an alternative consider now a class of $(n,k,\delta)$ convolutional
codes, i.e., a class of rate $k/n$ convolutional codes having
degree~$\delta$. We will demonstrate that for this class, the maximum
number of errors which can be corrected in some sliding window of
appropriate size is achieved by the subclass of MDP convolutional
codes.  In this paper, we will study the maximum number of erasures
that such a class of codes can decode and the conditions under which
this happens. Moreover we will show that over the erasure channel this
class of codes can decode extremely efficiently.

The paper is organized as follows. Section~\ref{Preliminaries}
provides the necessary background for the development of the
paper. Thus, subsection \ref{Erasure} explains the assumptions on
the channel model; subsection~\ref{MDP} provides all the
necessary concepts about MDP convolutional codes and their
characterizations. Section~\ref{Decoding} is the main part of the
paper. It contains our main result and describes in detail the
decoding procedure. It also provides examples and special
concerns to be noticed when comparing with MDS block codes, and
in particular with Reed-Solomon codes. Section~\ref{generatorM}
shows a  decoding method in which the transmitted information 
is recovered directly.

\section{Preliminaries} \label{Preliminaries}

\subsection{Erasure channel} \label{Erasure} 

An erasure channel is a communication channel where the symbols
sent either arrive correctly or the receiver knows that a symbol
has not been received or was received incorrectly. An important
example of an erasure channel is the Internet, where packet sizes
are upper bounded by 12,000 bits - the maximum that the Ethernet
protocol allows (that everyone uses at the user end). In many
cases, this maximum is actually used. Due to the nature of the
TCP part of the TCP/IP protocol stack, most sources need an
acknowledgment confirming that the packet has arrived at the
destination; these packets are only 320 bits long. So if everyone
were to use TCP/IP, the packet size distribution would be as
follows: 35\% --320 bits, 35\% -- 12,000 bits and 30\% -- in
between the two, uniform.  Real-time traffic used, e.g., in video
calling does not need an acknowledgment since that would take too
much time; overall, the following is a good assumption of the
packet size distribution: 30\% -- 320 bits, 50\% -- 12,000 bits,
20\% --in between, uniform.

We can model each packet as an element or sequence of elements
from a large alphabet. Since packets over the Internet are
usually protected by a cyclic redundancy check (CRC) code the
receiver knows when a packet is in error or has not arrived. For
the purpose of illustration we could employ as alphabet the
finite field $\F:=\F_{2^{1,000}}$. If a packet has less than
1,000 bits then one uses simply the corresponding element
of~$\F$. If the packet is larger one uses several alphabet
symbols to describe the packet. Even if one uses some
interleaving, such an encoding scheme results in the property that
errors tend to occur in bursts and this is a phenomena observed
about many channels modeled via the erasure channel. This point
is important to keep in mind when designing codes which are
capable of correcting many errors over the erasure channel.

\subsection{MDP convolutional codes}\label{MDP}
Let $\mathbb{F}$ be a finite field.  We view a convolutional code
$\mathcal{C}$ with rate $k/n$ as a submodule of
$\mathbb{F}^{n}[z]$ (see \cite{gl08,ro96a1,ro01}) that can
be described as
\[
\mathcal{C}=\left\{ \vec{v}(z) \in \mathbb{F}^{n}[z] |
  \vec{v}(z)=G(z)\vec{u}(z) \quad \text{with} \quad \vec{u}(z)
  \in \F^{k}[z] \right\}
\]
where $G(z)$ is a $n\times k$ polynomial matrix called a
\textbf{generator matrix} for $\mathcal{C}$, $\vec{u}(z)$ is the
\textbf{information vector} and $\vec{v}(z)$ is the \textbf{code
  vector} or \textbf{codeword}.

We define the \textbf{degree} of a convolutional code
$\mathcal{C}$, and we denote it by $\delta$, as the maximum of
the degrees of the determinants of the $k\times k$ sub-matrices of
any generator matrix of $\mathcal{C}$.  Then we say that
$\mathcal{C}$ is an $(n,k,\delta)$ convolutional code~\cite{mc98}.

In case the convolutional code $\C$ is also observable (see,
e.g., \cite{ro99a,ro01}) then $\C$ can be equivalently described
through a parity check matrix. In other words, there exists in
this case an $(n-k)\times n$ full rank polynomial matrix $H(z)$
such that
\[
\mathcal{C}=\left\{ \vec{v}(z) \in \mathbb{F}^{n}[z] \ \ | \ \ 
  H(z)\vec{v}(z)=\vec{0} \in \mathbb{F}^{n-k}[z] \right\}.
\]
 If we write $\vec{v}(z)=\vect{v}_{0}+\vect{v}_{1}z+\ldots+\vect{v}_{l}z^{l}$
(with $l\geq 0$) and we represent $H(z)$ as a matrix polynomial
\[
H(z)=H_{0}+H_{1}z+\ldots+H_{\nu}z^{\nu}.
\]
we can expand the kernel representation in the following way
\begin{eqnarray}\label{eq0}
\begin{sbmatrix}{ccccc}
  H_{0}   &   \\
  \vdots  & \ddots &  \\
  H_{\nu} & \ldots & H_{0}    \\
  & \ddots &         & \ddots  \\
  &        & H_{\nu} & \ldots & H_{0} \\
  &        &         & \ddots & \vdots \\
  &        &         &        & H_{\nu} \\
\end{sbmatrix}
\begin{sbmatrix}{c}
  \vect{v}_{0} \\
  \vect{v}_{1} \\
  \vdots \\
  \vect{v}_{l}
\end{sbmatrix}=\vec{0}.
\end{eqnarray}

An important distance measure  for convolutional
codes is the \textbf{free distance}:
\[
\dfree(\mathcal{C}):=\min \left\{ \wt(\vec{v}(z)) \ \ | \ \ 
  \vec{v}(z) \in \mathcal{C} \quad \text{and} \quad
  \vec{v}(z)\neq 0 \right\}.
\]
The following lemma shows the importance of the free distance as a
performance measure of a code used over the erasure channel.
\begin{lemma}
  If $\C$ is a convolutional code with free distance
  $d:=\dfree$ and if during transmission at most $d-1$ erasures
  occur then these erasures can be uniquely decoded. Moreover,
  there exist patterns of $d$ erasures which cannot be
  uniquely decoded.
\end{lemma}
\IEEEproof Let
$\vec{v}(z)=\vect{v}_{0}+\vect{v}_{1}z+\ldots+\vect{v}_{l}z^{l}$
be a received vector with $d-1$ symbols erased. Let the erasures
be in positions $i_1, \ldots, i_{d-1}$. The homogeneous
system~\eqref{eq0} of $(\nu+l+1)(n-k)$ equations with $(l+1)n$
unknowns can be changed into an equivalent nonhomogeneous system
$$
\hat H 
\begin{sbmatrix}{c}
  v_{i_1}\\
  v_{i_2}\\
  \vdots \\
  v_{i_{d-1}}
\end{sbmatrix}
=\vect{b} 
$$
of $(\nu+l+1)(n-k)$ equations with $d-1$ unknowns $v_{i_1},
\ldots, v_{i_{d-1}}$. 

This nonhomogeneous system has a solution, because of the assumption
that the channel allows only erasures. In addition the columns of the
system matrix are linearly independent, because
$d=\dfree(\mathcal{C})$, so the matrix $\hat H$ is full column
rank. It follows from these two facts that the solution must be unique. 
\endIEEEproof

Rosenthal and Smarandache~\cite{ro99a1} showed that an
$(n,k,\delta)$ convolutional code has a free distance upper bounded by
\begin{eqnarray}\label{eq1}
\dfree(\mathcal{C})\leq (n-k)\left(\left\lfloor 
\frac{\delta}{k}\right\rfloor + 1\right)+\delta +1.
\end{eqnarray}
This bound is known as the \textbf{generalized Singleton bound}
\cite{ro99a1} since it generalizes in a natural way the
Singleton bound for block codes.  Analogously, we say that an
$(n,k,\delta)$ code is a \textbf{maximum distance separable}
convolutional code (MDS)~\cite{ro99a1} if its free
distance achieves the generalized Singleton bound.

Another local distance measure, important as well for decoding
and related with the previous one, is the \textbf{column
  distance} \cite{jo99}, $d_{j}^{c}(\mathcal{C})$,
given by the expression
\[
d_{j}^{c}(\mathcal{C})=\min\left\{ \wt(\vect{v}_{[0,j]}(z))\ \ | \ \ 
  \vec{v}(z)\in \mathcal{C} \ \ \text{and} \ \ \vect{v}_{0}\neq
  0\right\}
\]
where $\vect{v}_{[0,j]}(z)=\vect{v}_{0}+\vect{v}_{1}z+\ldots+\vect{v}_{j}z^{j}$ represents the
$j$th truncation of the codeword $\vec{v}(z)\in \mathcal{C}$.  It is
related with the $\dfree(\mathcal{C})$ in the following way
\begin{eqnarray}\label{eq2}
\dfree(\mathcal{C})=\lim_{j\rightarrow \infty} d_{j}^{c}(\mathcal{C}).
\end{eqnarray}
The $j$-th column distance is then upper bounded by
\begin{eqnarray}\label{eq3}
d_{j}^{c}(\mathcal{C})\leq (n-k)(j+1)+1
\end{eqnarray}
and the maximality of any of the column distances implies the
maximality of all the previous ones, that is, if
$d_{j}^{c}(\mathcal{C})=(n-k)(j+1)+1$ for some $j$, then
$d_{i}^{c}(\mathcal{C})=(n-k)(i+1)+1$ for $i\leq j$, see
\cite{hu05,gl06}.  The $(m+1)$-tuple
$(d_{0}^{c}(\mathcal{C}),d_{1}^{c}(\mathcal{C}),\ldots,d_{m}^{c}(\mathcal{C}))$
is called the \textbf{column distance profile} of the code
\cite{jo99}.

Since no column distance can achieve a value greater than the
generalized Singleton bound, the largest integer for which that
bound can be attained is
\begin{eqnarray}\label{eq4}
L=\left\lfloor \frac{\delta}{k} \right\rfloor + 
\left\lfloor \frac{\delta}{n-k} \right\rfloor.
\end{eqnarray}
An $(n,k,\delta)$ convolutional code
$\mathcal{C}$ is \textbf{maximum distance profile} (MDP)
\cite{hu05,gl06}, if $d_{L}^{c}(\mathcal{C})=(n-k)(L+1)+1$.
In this case, every $d_{j}^{c}(\mathcal{C})$ for $j\leq L$ is
maximal, so we can say that the column distances of MDP codes
increase as rapidly as possible for as long as possible.

In order to characterize the column distances as well as MDP
codes  algebraically assume the parity check matrix is given as 
$H(z)=\sum^{\nu}_{i=0} H_{i}z^{i}$. For each  $j>\nu$ define
$H_j=0$ and define:
\begin{equation}                           \label{eq5}
\mathcal{H}_{j}= 
\begin{sbmatrix}{cccc}
  H_{0}  &         &        & \\
  H_{1}  & H_{0}   &        & \\
  \vdots & \vdots  & \ddots & \\
  H_{j} & H_{j-1} & \cdots & H_{0}
\end{sbmatrix}
\in \mathbb{F}^{(j+1)(n-k)\times(j+1)n}.
\end{equation}
Then we have:
\begin{theorem}(\cite[Proposition 2.1]{gl06})\label{Theorem-d}
Let $d\in\mathbb{N}$. Then the following properties are equivalent.
\begin{alphalist}
\item $d^c_j=d$;
\item none of the first~$n$ columns of~$\mathcal{H}_{j}$ is contained in
  the span of any other $d-2$ columns and one of the first~$n$
  columns of~$\mathcal{H}_{j}$ is in the span of some other $d-1$ columns
  of that matrix.
\end{alphalist}
\end{theorem}

As a consequence we have the algebraic characterization of MDP
convolutional codes:

\begin{theorem}(\cite[Theorem 3.1]{hu05})    \label{mdp}
The $j$-th column distance attains the maximum value
\begin{eqnarray}\label{eq4-1}
d_{j}^{c}=(n-k)(j+1)+1,
\end{eqnarray}
 if and only if, every
$(j+1)(n-k)\times(j+1)(n-k)$ full-size minor of $\mathcal{H}_{j}$ formed from the
columns with indices $1 \leq i_{1} < \cdots < i_{(j+1)(n-k)}$,
where $i_{s(n-k)}\leq sn$ for $s=1,\ldots,j$, is nonzero.

In particular when $j=L$, then $H(z)$ represents an MDP code, if
and only if, every $(L+1)(n-k)\times(L+1)(n-k)$ full-size minor of $\mathcal{H}_{L}$
formed from the columns with indices $1 \leq i_{1} < \cdots <
i_{(L+1)(n-k)}$, where $i_{s(n-k)}\leq sn$ for $s=1,\ldots,L$, is
nonzero.
\end{theorem} 

MDP convolutional codes can be thought to be like an MDS block code
within windows of size $(L+1)n$.  The nonsingular full-size minors
property given in the previous theorem ensures that if we truncate a
codeword at iterations up to $L$ it will have weight higher or equal than the
bound \eqref{eq4-1}.

\section{Decoding over an erasure channel}\label{Decoding}

Let us suppose that we use an MDP convolutional code
$\mathcal{C}$ to transmit over an erasure channel.
Then we can state the following result.
\begin{theorem}                \label{main}
  Let $\mathcal{C}$ be an $(n,k,\delta)$ MDP convolutional code.
  If in any sliding window of length $(L+1)n$ at most
  $(L+1)(n-k)$ erasures occur then we can recover the whole
  sequence.
\end{theorem}

\IEEEproof  Assume that we have
been able to correctly decode up to an instant $t-1$. Then we have the
following homogeneous system : {\small
\begin{eqnarray}\label{eqdec}
\begin{sbmatrix}{ccccccc}            
  H_{\nu} & H_{\nu-1} & \ldots & H_{0} \\
  & H_{\nu}   & \ldots & H_{1}   & \ddots \\
  &           &        &         &         \\
  &           & \ddots &         &         & H_{0} \\
  & & & H_{L} & \ldots & H_{1} & H_{0}
\end{sbmatrix}
\begin{sbmatrix}{c}
  \vect{v}_{t-\nu} \\
  \vdots \\
  \vect{v}_{t-1} \\
  \star \\
  \star\\
  \vdots \\
  \star
\end{sbmatrix}=0
\end{eqnarray}}
\normalsize where $\star$ takes the place of a vector that had some of
the components erased. Let the positions of the erased field elements be
$i_1, \ldots, i_{e},$ $e\leq (n-k)(L+1)$, where $i_1, \ldots,
i_{s},$ $s\leq n$, are the erasures occurring in the first erased
$n$-vector.  We can compute the syndrome and get a nonhomogeneous
system with $(L+1)(n-k)$ equations and $e$, at most $(L+1)(n-k)$,
variables.
 
We claim that there is an extension $\{ \vect{\tilde v}_{t},
\ldots,  \vect{\tilde v}_{t+L}\}$ such that the vector 
$ (\vect{v}_{t-\nu} ~ \ldots ~ \vect{v}_{t-1} ~ \vect{\tilde v}_{t},
\ldots,  \vect{\tilde v}_{t+L})$ is a
codeword and such that $\vect{\tilde v}_{t}$ is unique.

Indeed, we know that a solution of the system exists since we
assumed only erasures occur.  To prove the uniqueness of
$\vect{\tilde v}_{t}$, or equivalently, of the erased elements
$\tilde v_{i_1}, \ldots, \tilde v_{i_{s}},$ let us suppose there
exist two such good extensions $\{ \vect{\tilde v}_{t}, \ldots,
\vect{\tilde v}_{ t+L}\}$ and $\{ \vect{\tilde{\tilde v}}_{t},
\ldots, \vect{\tilde{\tilde v}}_{ t+L}\}$.  Let $\vect{h}_{i_1},
\ldots, \vect{h}_{i_e}$ , be the column vectors of the sliding
parity-check matrix in~\eqref{eqdec} which correspond to the
erasure elements. We have:
$$ \tilde
  v_{i_1} \vect{h}_{i_1}  + \ldots {\tilde
  v}_{i_{s}}\vect{h}_{i_s}+\ldots + \tilde
  v_{i_{e}}\vect{h}_{i_e}=\tilde{\vect{b}}$$  
and 
$$
\tilde{\tilde v}_{i_1} \vect{h}_{i_1} + \ldots +\tilde{\tilde
  v}_{i_{s}}\vect{h}_{i_s} + \ldots +\tilde{\tilde
  v}_{i_{e}}\vect{h}_{i_e}= \tilde{\tilde{\vect{b}}},$$
where the
vectors $\tilde{\vect{b}}$ and $\tilde{\tilde{\vect{b}}}$
correspond to the known part of the system.  Subtracting these
equations and observing that
$\tilde{\vect{b}}=\tilde{\tilde{\vect{b}}}$, we obtain: {\small
  $$
  (\tilde v_{i_1}-\tilde{\tilde v}_{i_1}) \vect{h}_{i_1} +
  \ldots +(\tilde v_{i_{s}}- \tilde{\tilde
    v}_{i_{s}})\vect{h}_{i_s}+\ldots + (\tilde v_{i_{e}}-
  \tilde{\tilde v}_{i_{e}})\vect{h}_{i_e}=0.
$$}
Using Theorem~\ref{Theorem-d} for a window of size $L$, and using that
the code is MDP, so $d^c_L=(L+1)(n-k)+1$, we obtain that, necessarily, 
\begin{align*} \tilde
v_{i_1}-&\tilde{\tilde v}_{i_1}=0,
~ \ldots,  ~\tilde
  v_{i_{s}}-
  \tilde{\tilde v}_{i_{s}}=0, \end{align*} by part (b) of
Theorem~\ref{Theorem-d}. This concludes the proof of our claim.

In order to find the value of this unique vector, we solve the  full column rank
system, find a solution and retain the part which is unique. Then we
slide $n$ bits to the next $n(L+1)$ window and proceed as above. 

\endIEEEproof

\subsection{Examples and Remarks}\label{example}

\begin{remark}
  The decoding algorithm requires only simple linear algebra. For
  every $(n-k)$ erasures a matrix of size at most $(L+1)(n-k)$
  has to be inverted over the base field $\F$. This is easily
  achieved even over fairly large fields.
\end{remark}

In addition one should notice that for a rate $\frac{k}{n}$ MDP
convolutional code, $100\cdot \frac{n-k}{n}$ percent of the erasures
can be corrected.

\begin{remark}
  Theorem~\ref{main} is optimal in a certain sense: One can show
  that for any $(n,k,\delta)$ code there exist patterns of
  $(L+2)(n-k)$ erasures in a sliding window of length $(L+2)n$
  which cannot be uniquely decoded.
\end{remark}

The following illustrative example compares the size of a
particular MDP convolutional code with an MDS block code which
would perform similarly.

\begin{example}
  Let us take a $(2,1,50)$ MDP convolutional code to decode over
  an erasure channel.  In this case the decoding can be completed
  if in any sliding window of length $202$ there are not more
  than $101$ erasures; $50\%$ of the erasures can be recovered.
  
  The MDS block code which achieves a comparable performance is a
  $[200,100]$ MDS block code.  In a block of $200$ symbols we can
  recover $100$ erasures, that is again $50\%$.
\end{example}

\begin{remark}\label{remark1}
  It has been noticed that the parameter $L$ gives us an upper
  bound on the length of the window we can take to correct, but
  it should be noticed as well that the property of
  Theorem~\ref{mdp} holds for every $j<L$.  This means that we
  can take smaller windows to set our systems (the size will be
  conveniently decided by the distribution of the erasures in the
  sequence).  Then in a window of size $(j+1)n$ symbols we can
  recover at most $(j+1)(n-k)$ erasures.
  
  This property allows us to recover the erasures in situations
  where the MDS block codes cannot do it.  For example, assume
  that we have been able to correctly decode up to an instant $t$
  and then it comes a block of $200$ symbols where $2$ bursts of
  $60$ erasures occur separated by a block of $80$ clean symbols, and
  after it, clean symbols again.
  \[
  \overbrace{\star \star \ldots \star \star}^{60}
  v_{61}v_{62}\ldots v_{140} \overbrace{\star \star \ldots \star
    \star}^{60}v_{201}v_{202}\ldots
  \]
  In this situation $120$ erasures happen in a block of $200$
  symbols and the MDS block code is not able to recover them.  In
  the block code situation one has to skip the whole block
  losing that information, and go on with the decoding.

   However, the MDP convolutional code can deal with this
  situation.  Let us set a $120$ symbols length window; in these
  windows we can correct up to $60$ erasures.  We can take $100$
  previous decoded symbols, then set a  window with the first $60$
  erasures and $60$ more clean symbols. In this way we can recover the
  first block of erasures.  Then we can slide through the
  received sequence with this $120$ symbols window until we set
  the rest of the erasures in the same way.
  \[
  v_{40}\ldots v_{140}\overbrace{\star \star \ldots \star
    \star}^{60}v_{201}v_{202}\ldots v_{260}  
  \]
  After this we have correctly decoded the sequence.
\end{remark}

\begin{remark}\label{remark2}
  Another advantage to remark is related to the storage and to
  the field size required to construct the codes.  In the example,
  we propose we have a $[200,100]$ MDS block code.  If we take,
  for example, a Reed-Solomon code (one of the most widely used
  MDS block codes) then we need to store the $200$ roots of a
  $200$ degree polynomial to set the code.  That is, we need at
  least $200$ field elements.
  
  However, to set the $(2,1,50)$ MDP convolutional code we need
  to store the coefficients of $2$ polynomials of degree $50$,
  that is at least $100$ different elements.
  
  Nevertheless there are some disadvantages.  On the one hand,
  the storage and the field size are smaller, but on the other
  hand, there are not direct constructions for the case of MDP
  convolutional codes.  This is still an open problem.
  
  The construction of MDP convolutional codes has been developed
  somewhat \cite{hu08}, however there exists still no efficient
  algorithm to construct this class of codes.  In relation to this
  problem, special type of matrices called \textit{superregular
    matrices} proved to be relevant during this study and this topic
  has become of main importance when trying to construct MDP
  convolutional codes \cite{hu05,ke06}.
  
  If we denote by $T^{i_{1},\ldots,i_{r}}_{j_{1},\ldots,j_{r}}$
  the $r\times r$ submatrix obtained from a matrix $T \in
  \mathbb{F}^{n\times n}$ by taking the rows with indices
  $i_{1},\ldots,i_{r}$ and the columns with indices
  $j_{1},\ldots,j_{r}$, then we can define a superregular matrix
  as follows.
  \begin{definition}\cite{hu05}
    A lower triangular Toepliz matrix $T$
    \begin{eqnarray}\label{eq6}
      T=\begin{sbmatrix}{cccc}
        t_{1}  &  0     & \ldots & 0      \\
        t_{2}  & t_{1}  & \ddots & \vdots \\
        \vdots & \ddots & \ddots & 0      \\
        t_{n}  & \ldots & t_{2}  & t_{1} 
      \end{sbmatrix} \in \mathbb{F}^{n\times n}
    \end{eqnarray}
    is said to be superregular if
    $T^{i_{1},\ldots,i_{r}}_{j_{1},\ldots,j_{r}}$ is nonsingular
    for all $1\leq r \leq n$ and all indices $1 \leq i_{1} <
    \ldots < i_{r} \leq n$, $1 \leq j_{1} < \ldots < j_{r} \leq
    n$ which satisfy $j_{s}\leq i_{s}$ for $s=1,\ldots,r$.  The
    submatrices obtained by picking such indices are called the
    \textit{proper submatrices} and their determinants the
    \textit{proper minors} of $T$.
  \end{definition}

  Unfortunately, the characterization or construction of these
  matrices is a hard problem and more research is needed in this
  direction in order to come up with a construction for MDP
  convolutional codes.
\end{remark}

\section{Decoding with the help of the generator matrix}  \label{generatorM}

In this section we explain how the use of the generator matrix of
the MDP code can make our decoding process more efficient and
faster.

We know that the encoding process is represented by
$G(z)\vec{u}(z)=\vec{v}(z)$, so the idea is to use this relation
to recover directly the original message $\vec{u}(z)$ instead of
computing first the code sequence and then decode it into the
original sequence $\vec{u}(z)$, as we did before when working
with the parity check matrix.

In an analogous way to the parity check matrix we can expand the
generator matrix into
\[
G(z)=G_{0}+G_{1}z+\ldots+G_{m}z^{m}
\]
and define $\mathcal{G}_{j}$ as
\begin{eqnarray}\label{eq7}
  \mathcal{G}_{j}=\begin{sbmatrix}{cccc}
    G_{0}  &         &        &        \\
    G_{1}  & G_{0}   &        &        \\
    \vdots &         & \ddots &        \\
    G_{j}  & G_{j-1} & \ldots & G_{0}
  \end{sbmatrix}.
\end{eqnarray}
Then the equivalences in the following theorem give us the
properties to improve the decoding algorithm.
\begin{theorem}(\cite[Theorem 2.4]{gl06}) Let $\mathcal{H}_{j}$
  and $\mathcal{G}_{j}$ be as in (\ref{eq5}) and (\ref{eq7}).
  Then the following are equivalent:
  \begin{enumerate}
  \item $d_{j}^{c}=(n-k)(j+1)+1$
  \item every $(j+1)k \times (j+1)k$ full-size minor of
    $\mathcal{G}^{T}_{j}$ formed from the columns with indices $1\leq
    t_{1}<\ldots<t_{(j+1)k}$, where $t_{sk+1}>sn$ for
    $s=1,\ldots,j$, is nonzero.
  \end{enumerate}
\end{theorem}
One notices that for an MDP convolutional code the maximum
size of the matrix $\mathcal{G}_{j}$ we can construct is again
given by the parameter $L$.  This tells us that the maximum
number of original symbols we are able to recover in one time is
$(L+1)k$.  Since in any sliding window of length $(L+1)n$ not
more than $(L+1)(n-k)$ erasures occur we can set a full rank
system with at least $(L+1)k$ equations to recover the $(L+1)k$
symbols of the original sequence. We will leave the details to
the reader.

\section{Conclusion}
In this paper, we propose MDP convolutional codes as an
alternative to block codes when decoding over an erasure channel.
We have seen that the step-by-step-MDS property of the MDP codes
lets us recover the maximum number of erasures at every step.
Even over large field sizes the complexity of decoding is
polynomial for a fixed window size since the decoding algorithm
requires the solving of some linear system only. Moreover, the sliding
window property allows us to adapt the decoding process to the
distribution of the erasures in the sequence.  We have shown how
the possibility of taking smaller windows lets us recover erasures
that the block codes cannot recover.

\section*{Acknowledgments}

We would like to thank Martin Haenggi who explained us the
distribution of packet sizes when transmitting files over the
Internet.

 \def\cprime{$'$} \def\cprime{$'$}

\end{document}